\documentclass[epj]{svjour}
\usepackage{graphics}
\usepackage{hyperref}
\hypersetup{colorlinks,allcolors=black}
\hypersetup{colorlinks = true, allcolors = blue}
\usepackage[utf8]{inputenc}
\usepackage{authblk}
\usepackage{hepparticles}
\usepackage{hepunits}
\usepackage{hepnames}

\begin{document}
\title{Hunt for rare processes and long-lived particles at FCC-ee}
\author{Marcin Chrzaszcz\inst{1} \and Rebeca Gonzalez Suarez\inst{2} \and St\'ephane Monteil\inst{3}
}                     
\institute{Henryk Niewodnicza\'nski Institute of Nuclear Physics (IFJ), Polish Academy of Sciences (PAN), Krak\'ow, Poland \and Uppsala University, Uppsala, Sweden \and Universit\'e Clermont Auvergne, CNRS/IN2P3, LPC, Clermont-Ferrand, France}
\date{Received: \today / Revised version: \today }
%
\abstract{
In this essay we discuss the possibilities and associated challenges concerning beyond the Standard Model searches at FCC-ee, such as rare decays of heavy-flavoured particles and long-lived particles. The Standard Model contains several suppression mechanisms, which cause a given group of processes to happen rarely, resulting in rare decays. The interest in these decays lies in the fact that the physics beyond the Standard Model does not need to be affected by the same suppression mechanism and therefore can naturally manifest in these decays. Their interest is reinforced by the recent report of several measurements of $b$-flavoured rare decays, showing deviations with respect to the Standard Model predictions. We will show how the FCC-ee project has unique capabilities to address these scientific questions and will consider the related detector design challenges to meet.   
Another group of processes discussed are those that produce new particles with relatively long lifetimes, that travel substantial distances inside the detectors before decaying. Models containing long-lived particles can give answers to many open questions of the Standard Model, such as the nature of dark matter, or the neutrino masses, among others; while providing an interesting experimental complement to mainstream searches. Long-lived particles often display unique experimental signatures, such as displaced tracks and vertices, ``disappearing'' tracks, or anomalously charged jets. Due to this, they are affected by very low background levels but in exchange, they often require dedicated reconstruction algorithms and triggers.  
The discovery of any of the discussed cases would have a critical impact in High Energy Physics and FCC-ee could provide a unique experimental opportunity to explore them. Moreover, the searches proposed here could motivate an out-of-the-box optimization of the experimental conditions, that could bring in innovative solutions, such as new, possibly very large  tracking  detectors; or cutting-edge reconstruction algorithms  that would boost the FCC-ee reach for unusual final states. 
\PACS{
      {PACS-key}{describing text of that key}   \and
      {PACS-key}{describing text of that key}
     } 
} 
\maketitle

\section{Introduction} \label{section:intro}
The future FCC-ee is a frontier Higgs, Top, Electroweak, and Flavour factory. It will be operated in a 100\,km circular tunnel built in the CERN area, and will serve as the first step of the Future Circular Collider (FCC) integrated program towards $\geq$ 100\,TeV proton-proton collisions in the same infrastructure~\cite{CERN-ESU-015}. 

In addition to an essential and unique program of high-precision Standard Model (SM) measurements, FCC-ee offers powerful opportunities for the discovery of new phenomena. Direct and indirect evidence for physics beyond the Standard Model (BSM), can be achieved at FCC-ee via a combination of high precision measurements, where new physics could manifest as small deviations with respect to the SM predictions; and also via searches, both focusing on specific BSM models, or in a model-independent way, targeting final state signatures or sensitive areas of the phase space. 

In this essay two broad lines are proposed and highlighted for dedicated exploration at FCC-ee: rare and suppressed decays, and long-lived particles that could give rise to distinct physics signatures.

\section{Rare and suppressed decays of heavy-flavoured particles} \label{section:flavour}

\subsection{Introduction to rare heavy-flavoured decays}
The ensemble of rare decays of heavy-flavoured particles is formed by Flavour-changing neutral current (FCNC) decays with either photons or leptons in the final state. These decays proceed in the SM through electroweak penguins and/or box loop diagrams involving heavy SM particles ($W$, $H$, top quark) that are further suppressed by either non-diagonal CKM matrix elements (for $b$-flavoured particles)  
and GIM mechanism (for $c$-flavoured particles). There are a number of BSM scenarios where neither the loop suppression nor the CKM and GIM suppression occur. The experimental study of those decays benefits from clear detection signatures. Combined with the large statistics of heavy-flavoured particles produced at FCC-ee, this kind of process will definitely continue to be an excellent laboratory to search for BSM physics. The measurement of the branching fraction of $b \to s \gamma$ at LEP and asymmetric B-factories for instance, is still one the most significant constraints of the parameters of supersymmetric models~\cite{Arbey:2012ax}, as it is the more recent measurement of the $B_s \to \mu^+\mu^-$  branching fraction by the LHC experiments~\cite{Aaij:2017vad}. The current experimental knowledge concerns the final states with photons or light leptons. The final states comprising $\tau$ particles are not yet observed and their search is definitely a must do at FCC-ee.    

The study of rare $b$-flavoured hadron decays by the LHCb experiment have brought a number of departures between measurements and SM predictions. Though none of them is significant enough to provide a claim that the SM is defeated, the departures add up consistently to provide a direction towards new Physics scenarios. To that extent, the transitions involving third generation fermions $b \to s \tau \tau$ is of particular interest since it can sort out the possible classes of models. 

The measurements of the heavy-flavoured rare decays with $\tau$ leptons open unique physics opportunities at FCC-ee; however, contrarily to the final states with light leptons, the experimental signatures involve neutrinos and the experimental reconstruction imposes challenges to the detector design that this essay will approach to sketch. 

The rare decays that are allowed in the SM are discussed in a companion essay of this volume~\cite{Monteil:2021ith}. This section focuses instead on the forbidden modes in the SM in which the lepton number conservation is violated. Final states with $\tau$ leptons are selected for this exploration, where FCC-ee provides unique opportunities.

\subsection{Partial reconstruction techniques}
The decays of $b$-flavoured hadrons containing $\tau$ particles are characterized by at least a neutrino in the final state. The inability to efficiently detect them in these rare decays makes most of the experimental search challenging. However, an excellent knowledge of the production and decay vertices of the parent $b$-flavoured particle can allow to to fully solve the kinematics of these decays and provide a complete reconstruction of the invariant mass of the candidates. The consideration of the multibody hadronic $\tau$ decays can be further used to improve the knowledge of the decay chain and improve the invariant-mass resolutions. It is possible to think of an example with one neutrino in the final state originating for a charged three-body $\tau$ decay. Namely, the measurement of the primary vertex and the decay vertex of the $b$-flavoured hadron provides an estimate of the momentum direction of the parent particle, hence fixing two unknowns. The reconstruction of the $\tau$ lepton vertex, allowing to infer its momentum direction, adds two constraints and fixes two additional degrees of freedom. The constraints are however not fully independent and the kinematic of the decay is solved up to a quadratic ambiguity. The knowledge of the mass of the $\tau$ lepton can be further used to close the system. Supplementary requirements are in order if there are more neutrinos in the final state. 

Experiments at FCC-ee potentially provide unique characteristics to perform this type of vertex-constrained kinematical fit. The $b$-quark fragmentation functions at the $Z$ pole have been empirically determined at LEP experiments~\cite{Heister:2001jg,DELPHI:2011aa} to be hard. The average energy of $b$-flavoured particles produced in $Z \to b \overline{b}$ decays is about 35 GeV. The average decay length of these particles is in turn about 3~mm. The production and decay vertices of these decay lengths can be resolved with high precision with tracking detectors of few micrometers hit space resolution. The controlled rate of beam-induced backgrounds at FCC-ee allows, in principle, to place these detectors close to the interaction point, providing an adequate leverage to perform the pointing extrapolation of tracks. Furthermore, the distribution as a function of the polar angle $\theta$ of the $b$-flavoured particles, governed by the underlying vectorial and parity-violating axial-vector couplings of the $Z$ to $b$ quarks evolves as 

\begin{equation}
 \frac{{\rm d} \sigma(B)}{{\rm d \cos \theta}} \propto {\cal A}_0 (1 +\cos^2 \theta) + {\cal A}_1 \cos \theta, 
\end{equation}  

\noindent where ${\cal A}_0$ and ${\cal A}_1$ stand for normalisation factors (${\cal A}_1$ embodying the parity-violating asymmetry) and $B$ stands for a $b$-flavoured state produced at the $Z$ pole. Most of the $b$-flavoured particles are hence contained in the barrel of the vertex detector.      

The yet unobserved $B^0 \to  K^*(892)^0  \tau^+ \tau^-$ decay has been studied in the context of the FCC-ee conceptual design report~\cite{Abada:2019lih} by means of Monte Carlo simulated events analysed with a parametric detector. The performance of the vertex detector considered in the study is inspired by that obtained for the ILD vertex detector design~\cite{Behnke:2013lya}. The same approach is followed in this essay to characterize the Lepton-Flavour Violating (LFV) $b$-flavoured particle decays. The study of the  $B^0 \to  K^*(892)^0 \tau^+ \tau^-$ decay is described in more details in the companion essay in this volume~\cite{Monteil:2021ith}.

\subsection{Lepton-Flavour Violating decays of $b$-flavoured particles.}
Among the class of dileptonic studies in $b \to s,d$  quark transitions, the measurements of the decays $H_b \to H_{\rm light}  \tau \ell$, where $H_f$ denotes a hadron of flavour $f$ and $\ell$ an electron or a muon, are of high interest as indisputable evidence for physics beyond the SM. They are present as an illustration in most of the models inspired by the Flavour anomalies~\cite{Cornella:2021sby,Hiller,Becirevic:2017jtw,Altmannshofer:2017yso}. They are contributing beyond these realisations in the understanding of flavour patterns that might relate the third generation of quarks and leptons. A recent measurement performed by the LHCb experiment~\cite{Aaij:2020mqb} has first settled the experimental scene by putting an upper limit on the branching ${\cal B}(B^+ \to K \mu^-\tau^+)$ at the level of few $10^{-5}$ at 90 \% C.L. (whereas limits on two light flavours in the final states were already obtained by $B$-factories, {\it e.g.}~\cite{Sandilya:2018pop}, at the level of a few $10^{-7}$). Topological reconstruction techniques similar to those advocated above these lines were used. 

\begin{figure*} 
\centering
\resizebox{0.45\textwidth}{!}{\includegraphics{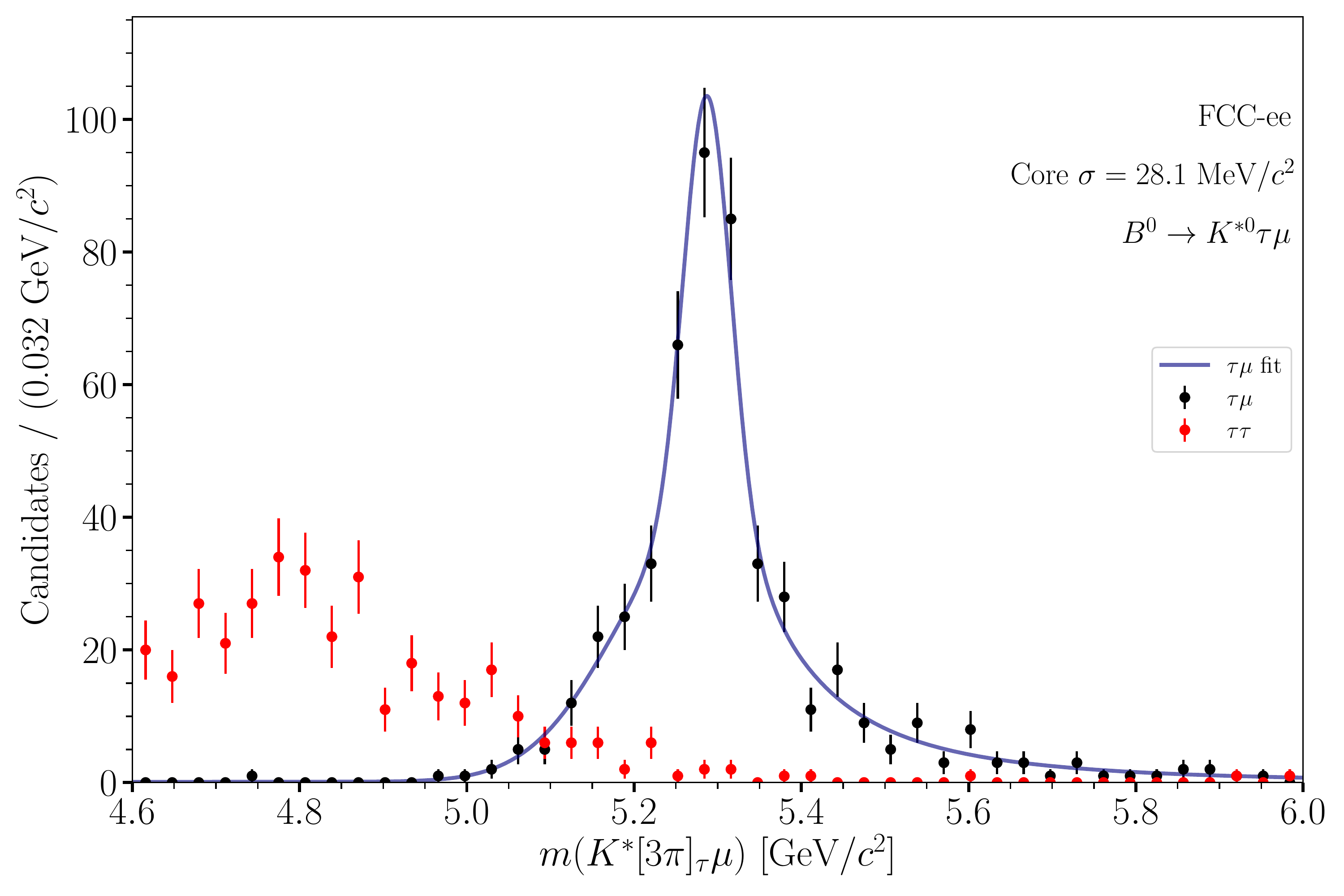}}
\caption{Invariant-mass distribution of $B^0 \to  K^*(892)^0  \tau \mu$ candidates, determined thanks to the constraints brought by the measurements of the decay lengths of the intermediate particles of the decay. Events are simulated using a fast simulation featuring the IDEA detector concept momentum resolution~\cite{Abada:2019zxq}. The resolution on the decay vertices are simulated as described in the text. Simulated events containing $B^0 \to  K^*(892)^0  \tau^+ \tau^-$ are as well reconstructed with the same topological method and superimposed. A model made with two Crystal-Ball functions   and a gaussian is fit to the signal events. The core resolution obtained from the gaussian width fit is displayed on the figure.}
\label{fig:KstarTauMu}      
\end{figure*}

A fast parametric simulation featuring momentum reconstruction resolutions as of~\cite{Abada:2019zxq} and emulating a high resolution vertex detector performance is used to simulate LFV signal events $(B^0 \to K^{*0} \mu^-\tau^+)$. As an illustration of the requirements that can be placed on the detector, the figure~\ref{fig:KstarTauMu} displays the invariant-mass distribution of the events reconstructed by means of the topological reconstruction technique, where the secondary and tertiary vertices are known with a resolution of $10$ and $50$ $\mu{\rm m}$ in the transverse and longitudinal directions,  with respect to the flight paths of each decaying particles. The simulated mass of the $B^0$ particle is recovered with a typical resolution of $30$ MeV/$c^2$. The reconstructed invariant-mass of $B^0 \to  K^*(892)^0  \tau^+ \tau^-$ decays events, where one of the $\tau$ particle decays into a muon (hence mimicking the LFV final state) is superimposed, illustrating the potential separation power of the topological reconstruction method. The number of LFV events gathered in the plot corresponds to a branching fraction of ${\cal O}(10^{-7}$ and is scaled to the baseline luminosity expected at FCC-ee. An equivalent number of $B^0 \to  K^*(892)^0  \tau^+ \tau^-$ is considered for the sake of this illustration. Though the actual study including realistic background sources is to be performed, this exploratory study illustrates the possible requirements on the design and key performances of the vertex detector that will be needed to be achieved.           

\section{The lifetime frontier} 
The known particles of the SM have different lifetimes that range from the very short, decaying right after being produced in collisions like the Higgs boson ($10^{-22}$~s); to the very large, like the proton, on the order of $10^{34}$ years and above. New BSM particles, can also display a variety of lifetimes, related to their mass and couplings among other parameters. 

When colliders push the energy frontier, heavier and shorter-lived particles become accessible, and detector design, trigger and reconstruction techniques reflect this fact. Mainstream searches for new particles in colliders therefore focus on prompt new particles decaying right after being produced. However, particles that travel substantial distances within the detector before decaying should also be considered, and FCC-ee is no exception. 

Long-lived particles (LLPs) are strongly motivated theoretically, appearing in many different scenarios that could shed light on important open questions of particle physics, such as dark matter (DM), neutrino masses, or the Baryon Asymmetry of the Universe (BAU), to name a few. In this essay we will briefly discuss some of the most striking possibilities for this kind of searches at the FCC-ee.

One of the most attractive aspects of LLPs relates to their uncommon experimental collider signatures. Some LLPs decay after flying some distance from the primary interaction point, producing displaced vertices, with decay products including charged and neutral SM particles (e.g. charged leptons and pions). Other BSM models predict disappearing LLPs giving rise to ``short'' or ``broken'' tracks; some LLPs are ``stopped'' or delayed in time; other LLP produce unusual jets, such as track less-jets and ``dark showers''. Such experimental signatures are distinct from SM processes and each of them would, if observed, constitute a smoking gun of new physics. 

Although LLP signatures offer great potential for searches and are affected by very low, sometimes only instrumental, background, standard reconstruction techniques at colliders are often unable to identify them and they are typically very hard to trigger on. 

The future FCC-ee offers exciting potential for the study of LLP, where searches can be not only complementary to similar searches at collider and non-collider experiments, but highly competitive. The large integrated luminosity of the FCC run around the Z pole, producing 5 $10^{12}$ $Z$ bosons~\cite{Benedikt:2651299}, known as the Tera-Z regime, will allow for example for the direct search of new, feebly interacting particles that could be either sound DM candidates, or closely linked to neutrino masses and the BAU (or both) and manifest long-lived signatures~\cite{Agrawal:2021dbo}. Such are Axion-like Particles (ALPs) or Heavy Neutral Leptons (NHL) and their subcategories: Sterile or Right-Handed Neutrinos.

Concerning HNL searches, electron-positron colliders, such as FCC-ee are expected to provide the best sensitivity to low neutrino mixing angles via displaced vertex  searches~\cite{doi:10.1142/S0217751X17500786}. The large statistics Tera-Z run is expected to be particularly powerful in this area~\cite{Blondel:2014bra,Abada:2019zxq,Klaric:2020lov,Abada:2018oly}.

A preliminary study was done for the benchmark case of HNL produced in the process $Z\rightarrow\nu N$ that subsequently decay into a $W$ boson and a lepton with a $N \rightarrow lq\bar{q}$ final state~\cite{Blondel:2014bra}. For low values of the neutrino mixing angle, the decay length of the heavy neutrino can be significant The results showed that the Tera-Z run would allow for sensitivity down to a heavy-light mixing of $10^{-11}$, covering a large phase-space for heavy neutrino masses between the $b$-hadron mass scale (5~GeV) and the electroweak mass scale (80~GeV) with displaced vertex searches. It is important to note that this kind of searches are affected by very little background, since the displacement of the vertex can be e.g. of 1~m. Sufficiently long-lived HNL could also potentially allow for neutrino oscillations to be studied~\cite{Antusch:2017ebe}.

Hidden sectors, that propose parallel, non-SM, collections of particles and forces, that interact with the SM very feebly are another compelling solution to different SM problems that results in LLPs. For example, the possibility of a new, Dark Sector connected with DM, stands out as a good alternative to classic DM models.

Dark Sectors could be similar to the SM, containing DM states, distinct force carriers and matter fields, and dark Higgs bosons. New Dark Sector particles can decay in different ways: promptly, with a displaced vertex, or outside the detector if they are collider stable; and all three regimes need to be probed. The future FCC-ee has the potential to unveil new Dark Sectors via LLPs~\cite{Abada:2019lih}.

The evidence for a Dark Sector could for example come in the form of ALPs, pseudo-scalar particles generally very weakly-interacting, that are predicted by many extensions the SM, most notably string theory. At lepton colliders, ALP production in association with a photon, a $Z$ boson or a Higgs boson provide the dominant production processes. For small couplings and light ALPs, the ALP decay vertex can be considerably displaced from the production vertex. Very long-lived ALPs would leave the detector before decaying, leaving a trace of missing energy. As it has been shown in~\cite{Bauer:2018uxu}, a high-luminosity run at the $Z$ pole would significantly increase the sensitivity to ALPs produced in $e^+e^-\rightarrow \gamma a$ with subsequent decays $a\rightarrow \gamma\gamma$ or $a \rightarrow l^+l^-$. 

The hierarchy problem is another well-known SM problem. Twin Higgs models, similar in nature to composite Higgs models, provide a solution to it. In some models, an entire Twin copy of the SM is proposed, with many new light states that the SM Higgs boson can decay to. Collider signatures for this class of models include displaced exotic Higgs boson decays, with decays length depending on the interaction strength. At FCC-ee the most relevant regions of parameter space in these models are experimentally accessible~\cite{Abada:2019lih}. 

Other BSM models producing exotic, long-lived Higgs decays that FCC-ee could be sensitive to, include for example, Hidden Valley scenarios. In some Hidden Valley models a new sector, weakly coupled to the SM results in neutral long-lived particles that the Higgs boson can decay to~\cite{Alipour-Fard:2018lsf}.

The opportunities do not end here, other options that could also be exploited at FCC-ee are detector-stable and massive LLPs including fractionally-charged particles, or even inflatons~\cite{Bezrukov:2013fca}. It is worth to evaluate all the different possibilities in the near future to be able to strategically optimize FCC-ee towards discovery.

\section{Outlook and conclusions} 
In order to explore the options presented in this essay, detector challenges will have to be met. 

The rare decays of $b$-flavoured hadrons with $\tau$ leptons in the final state in general and the lepton-flavour-violating decays in particular pose demanding requirements to the vertex finding detectors. The reconstruction of the undetected tauonic neutrino relies on the reconstruction of the production and decay vertices of the parent $b$-hadron. The performance sketched by the detector developments of the International Linear Collider (ILC), a proposed linear $e^+e^-$ collider in Japan~\cite{Behnke:2013lya}, would need to be enhanced to allow for these measurements. It is instrumental that improvements in vertex location resolution are examined in the context of FCC-ee: the distance of the first active layer from the interaction point (and hence the constraints on the beam pipe design) must be addressed, the impact of the hit space resolution and the detector alignment constraints are to be devised, novel technology and geometry such as bending silicon pixels, shall be considered. It is expected that the search sensitivity scales linearly with the invariant-mass resolution resolution, which in turn depends on the decay vertex position precision. These dependencies will have to be exercised and quantified with actual detector simulations including realistic background reconstructions.      

Creative work will be in order to achieve ultimate sensitivity in searches for LLPs at FCC-ee. Luckily, FCC-ee offers many experimental opportunities. First of all, it is possible to envisage up to four detectors at FCC-ee, two of which sitting in the very large caverns required from the start to be used by the subsequent hadron collider detectors. The caverns are foreseen to be deep underground, up to 300 meters, reducing considerably the cosmic ray backgrounds. A detector fully optimized for longer lifetimes can thus be considered, taking into account economic constrains.  
    
For conventional general purpose detectors, optimization studies to detect unusual experimental signatures could bring in innovative solutions, e.g. new, possibly relatively large-sized tracking detectors, or associated algorithms that would be optimal for these type of searches in unusual final states. 

\begin{figure*} 
\centering
\resizebox{0.45\textwidth}{!}{\includegraphics{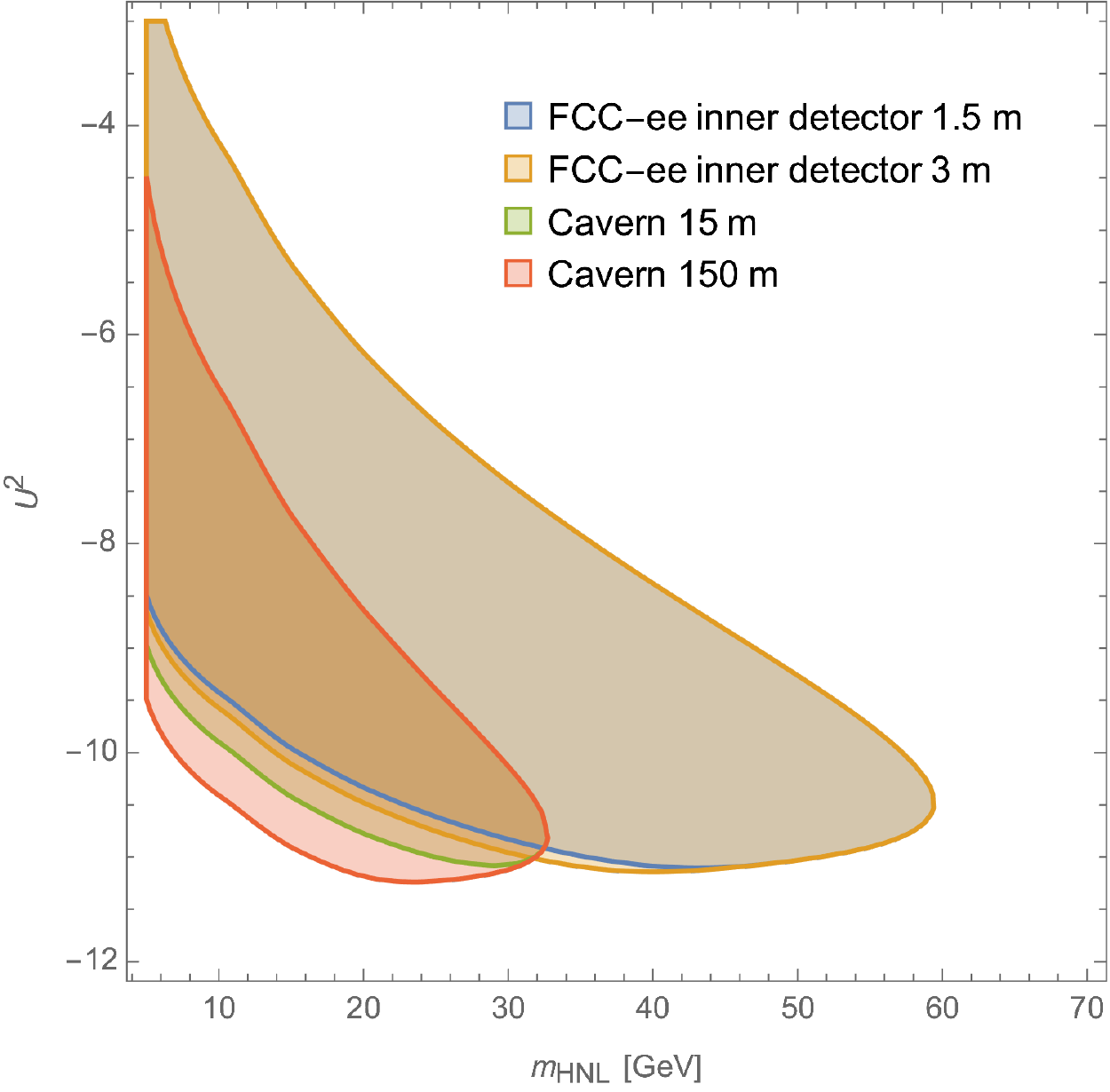}}
\caption{Expected exclusion region of HNL at FCC-ee as a function of the mass and neutrino mixing. The blue area corresponds to the expected sensitivity of a generic FCC-ee detector, not optimized for LLPs with a radius of 1.5~m, the orange area shows the effect of doubling that radius. The green and red areas illustrate the sensitivity for HNL that the decay in a typical 15~m cavern and an unrealistic one of a 150~m radius respectively.}
\label{fig:llp}      
\end{figure*}

To give an example, for the flagship HNL searches there is potential gain when increasing the experimental volume. A typical FCC-ee detector can potentially reconstruct HNL from $1\rm cm$ to $1.5\rm m$ in an essentially background free environment assuming a HNL mass over $5 \rm GeV$. The sensitivity of that baseline, together with three alternative scenarios is considered and shown in Fig.~\ref{fig:llp}. The blue and orange areas correspond to an inner detector size of 1.5, about standard, and 3~m respectively. Doubling the size of the tracker would clearly increase the reach for low mixing angle for HNL of masses 40~GeV and below. Then it is possible to see the further improvement for low mixing angles and masses of 30~GeV and below that can be achieved by equipping the cavern with dedicated instrumentation in the spirit of~\cite{Curtin:2018mvb}, assuming a typical cavern size of 15~m of radius, in green. For comparison, it is also possible to show the effect on the HNL searches at FCC-ee when instrumenting an unrealistic cavern of $150\rm m$ of radius, shown in the same plot in red. It is clear that the cavern offers a good compromise in terms of sensitivity without the need for expensive civil engineering~\cite{Chrzaszcz:2020emg} with about half one order of magnitude of potential gain on the $U^2$ mixing parameter .

Timing will also be essential in detector design for FCC-ee to be sensitive to $\beta < 1$, heavy particles leading to out-of time or even stopped decays. The ability to infer the LLP mass from its time-of-flight and the $e^+e^-$ constrained event kinematics should be exploited and experimental solutions such as timing layers in the tracker, explored. The data acquisition system should adapt to the continuous beam crossing rate of ~50MHz, given the essentially off-time characteristics and the absence of signals in the inner detectors in many  signatures. 

Finally, it is worth to discuss that when establishing the FCC tunnel and related civil engineering that will be necessary for its physics program it is possible to plan towards maximizing the physics reach by preparing supplementary tunnels or experimental areas to house additional experiments. 

In this essay we have suggested and discussed two broad BSM searches to pursue and exploit at the future FCC-ee: rare and suppressed decays, and long-lived signatures. The observation of any of the discussed scenarios would potentially shape the future of High Energy Physics, and they are just examples of what the FCC-ee could be sensitive to beyond precision measurements. Furthermore, innovative experimental solutions will be needed in order to fully exploit the potential of the searches proposed in this essay, boosting the reach of the FCC-ee for non-standard signals.

\bibliographystyle{jhep}
\bibliography{references}
\end{document}